\documentclass[aps,prl,showpacs,preprint]{revtex4}
\usepackage{graphics}
\usepackage[dvips]{color}
\usepackage[T1]{fontenc}
\usepackage[latin1]{inputenc}
\usepackage{graphicx}
\usepackage{amssymb}
\usepackage{rotating}
\usepackage{epsfig}

\input epsf.sty

\newcommand{\ie}{{\em i.e.}}
\newcommand{\eg}{{\em e.g.}}

\begin{document}

\title{
Collapse transition of flexible homopolymers with adhesive contacts}
\author{Sid Ahmed Sabeur}
\author{Mounira Bouarkat}
\affiliation{D\'epartement de Physique, Facult\'e des Sciences, USTO, Oran 31000, Algeria
}
\author{Friederike Schmid}
\affiliation{Institut f\"ur Physik, Johannes Gutenberg--Universit\"at, Staudinger Weg 7, D--55099 Mainz, Germany}

\date{\today}

\begin{abstract}

The presence of ordered structures such as helices in collapsed states of
polymer chains is still an open question challenging physics and biology.
In this work, we present a potential model for polymer chains with
monomers that are not strictly attractive, but that can make adhesive
contacts with other monomers. We find that the chain develop helical order
during the process of collapsing from an initially stretched
conformation. It seems in this case that the adhesive contacts help the
polymer chain to stay trapped in the helix state.

\end{abstract}


\maketitle

\section{Introduction}
The three dimensional structure of a polymer chain
will determine many of its properties. The most important examples of polymers
are synthetic homopolymers, such as polyethylene an biopolymers, like proteins and DNA.
In the case of synthetic polymers, the structure is the basis of a multitude
of macroscopic products of everyday life\cite{Vogel}.
On the other hand, biopolymers are the basic building blocks of cells and
involved in the majority of biological processes such as chemical reaction catalysis, oxygen transport
or muscles contractions\cite{Poulain}. The ground states of these molecules are
often crucial to their functions.

In recent years, designing functional polymers at nanometers scales has gained increasing interest\cite{Glotzer}.
However, it is challenging to control structures that have the ability to undergo cooperative transitions between
random conformations (globules) and ordered conformations ($\alpha-$helices and $\beta-$sheets). There has been a spate of theoretical studies
dealing with this issue, commonly achieved using Monte Carlo and molecular dynamics simulations\cite{Frisch2}. These works have offered rewarding insights for understanding the dynamics and the phase transition of polymers and there is still a big interest in investigating them until today\cite{Landau5}.

In a previous work, we have found that flexible homopolymers spontaneously develop helical order during the process
of collapsing from an initially stretched conformation. We have also demonstrated that the helices are long lived transient states at low temperatures\cite{Sabeur1}, but with "borrowing" energy from their surrounding solvent particles, they overcome the energetic barrier and collapse into a stable globule.

Here, we extend our study by adding to the polymer model adhesive contacts between non covalently linked monomers.
These kind of contacts appears for example in the vital part of the inflammatory response, where the activated cells, due to adhesive contacts develop capabilities to transmigrate into the inflamed tissue\cite{Heinrich}.

The report is organized as follows. In Sec. II we describe the model, simulation method
and the parameters used to quantify the ordered structures. In Sec. III, we present the results from the simulations.
Finally, we conclude with a summary of our results and the appropriate acknowledgments.

\section{Model and Methods}
In the simulation, we have used a simple bead-spring model. The polymer chain consists of $N$
identical monomers where each monomer position varies continuously in three dimensions.
Dimensionless units are used during the simulation and are defined in terms
of the bead size $\sigma$ and the Lennard-Jones energy $\epsilon$. There are three types
of interactions: a harmonic bond length interaction between adjacent monomers

\begin{equation}
\label{eq:vbond}
 V_{\mbox{\tiny bond}}(r)=a(r-r_0)^2,
\end{equation}

with $a=100\epsilon/\sigma^2$ and $r_0=0.85\sigma$

A Lennard-Jones potential between non adjacent monomers which is
repulsive up to a distance $x=\sigma$.

\begin{equation}
\label{eq:vrep}
V_{\mbox{\tiny rep}}(r)=\left\{
\begin{array}{l}
[(\sigma/r)^{12}-2(\sigma/r)^6+1 ]
\;\mbox{for $r<\sigma$}\\
0\quad \mbox{otherwise},
\end{array}
\right.
\end{equation}

In this potential, we have removed the attractive part but the
excluded volume part is kept and adhesive interactions between
any two monomers along the chain are possible. These interactions have
three properties:

Each monomer can only have one adhesive bond. \ie, if a monomer establish an adhesive contact with another monomer along the chain,
establishing adhesive contacts with other monomers will be prohibited. This is important because it makes
all the difference between the adhesive bond and general attractive interactions.
To get collapse, the minimum energy of the bond should be negative. To
study unraveling dynamics of the polymer chain, there should be an energy
barrier between the adhesively bonded and non bonded state.

Assuming the validity of the Kramer's rate theory\cite{Kramer}, we
have introduced an additional variable $\chi_{ij}$ that describes
the state between  two monomers i and j.

If $\chi_{ij}=1$, there is an adhesive bond , if $\chi_{ij}=0$,
there is none.

The interaction potential for non-bonded monomers is then:

\begin{equation}
V_{\mbox{\tiny inter}}(r)=\sum_{ij}[V_{\mbox{\tiny
rep}}(r_{ij})+\chi_{ij} V_{\mbox{\tiny adbond}}(r_{ij})]
\end{equation}
with

\begin{equation}
\label{eq:vadbond}
V_{\mbox{\tiny adbond}}(r)=\left\{
\begin{array}{l}
k((r/\sigma)^2-1.)-\varepsilon \quad \mbox{for $r<2.5 \sigma$}\\
V_{0}\quad \mbox{otherwise},\\
\end{array}
\right.
\end{equation}

This potential is constructed such that the minimum value of
$V_{\mbox{inter}}(r)$ is $-\varepsilon$ in the case of bonded
monomers (zero for non-bonded). Hence, an adhesive bond has the
energy $-\varepsilon$. $V_{0}=V_{\mbox{\tiny inter}}(2.5\sigma)$ is
the height of the energy barrier, thus the cutoff $r_{\mbox{\tiny
cut}}$ should be chosen large enough that $V_{0}$ is positive. Bonds
can only be established between monomers with distances less than
$r_{\mbox{\tiny cut}}$ and may not be established between
monomers that are neighbors along the chain (monomers $n$ and
$n+1$), \ie, $\chi_{n,n+1}=0$.

In addition, there is a constraint $\sum_j \chi_{ij}\leq 1$, \ie,
every monomer can have only one bond.

\begin{figure}[t]
\includegraphics[scale=0.5,angle=0]{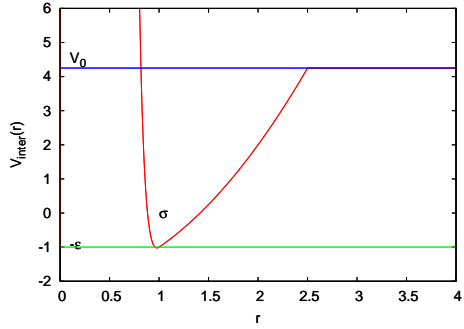}
\vspace*{-0.3cm} \caption{\label{fig1} The non-bonded interaction
potential}
\end{figure}

In simulating the dynamics of this model, We have used a mixture of
Monte Carlo for bond-breaking/bond-establishing and Langevin
dynamics. One Monte Carlo step consists of enumerating all monomer
pairs which are not bonded along the chain, than picking randomly
one pair $(i,j)$, and if $\chi_{ij}=1$, break adhesive bond with
probability $\exp(-\beta(V_{0}-V(r_{ij})))$. If $\chi_{ij}=0$ and
the bond is not forbidden by any constraint, establish adhesive bond
with probability $-\beta V_{0}$, than repeat this a number of times
(\eg, Ntimes, once for each monomer). Langevin Dynamics step without
hydrodynamics describes the motion of the monomers.

\subsection{Order parameter}
To characterize the helical structures, we use an order parameter that is commonly
used to represent the net helical growth in chainlike molecules\cite{Kemp2}, as
defined by

\begin{equation}
H_4=\left( \frac{1}{N-2}\sum_{i=2}^{N-1}\textbf{u}_i\right)^2
\end{equation}

\begin{figure}[t]
\includegraphics[scale=0.7,angle=0]{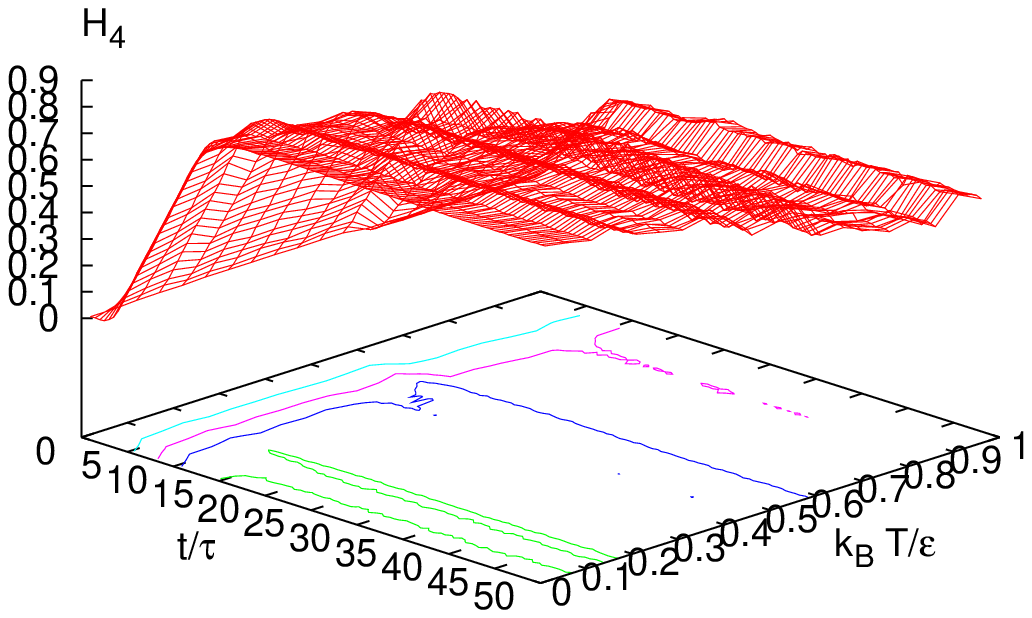}
\end{figure}

\begin{figure}[t]
\includegraphics[scale=0.7,angle=0]{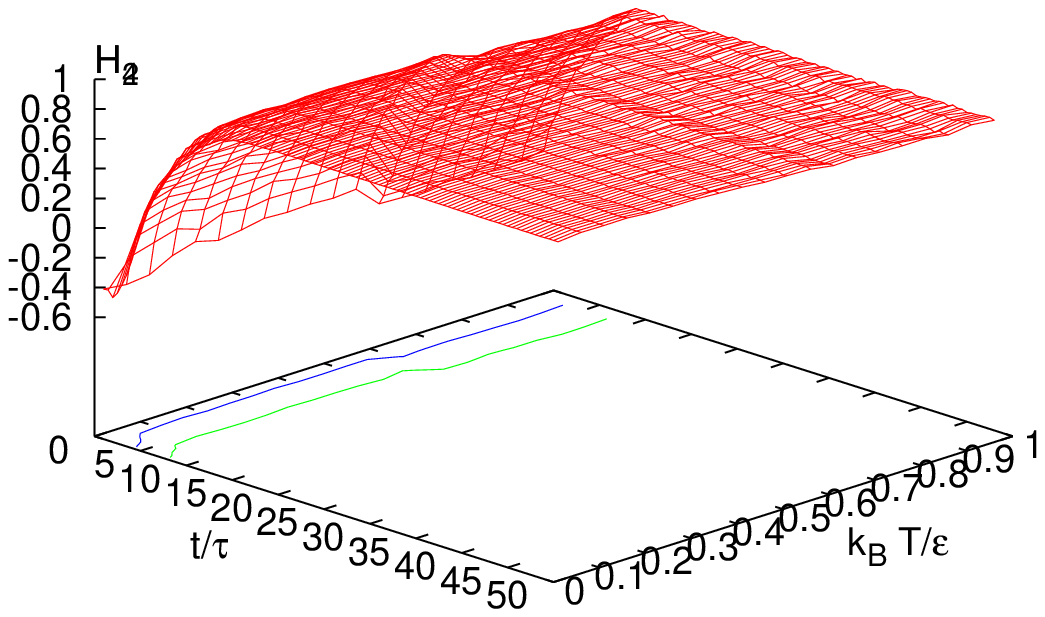}
\end{figure}

\begin{figure}[t]
\includegraphics[scale=0.7,angle=0]{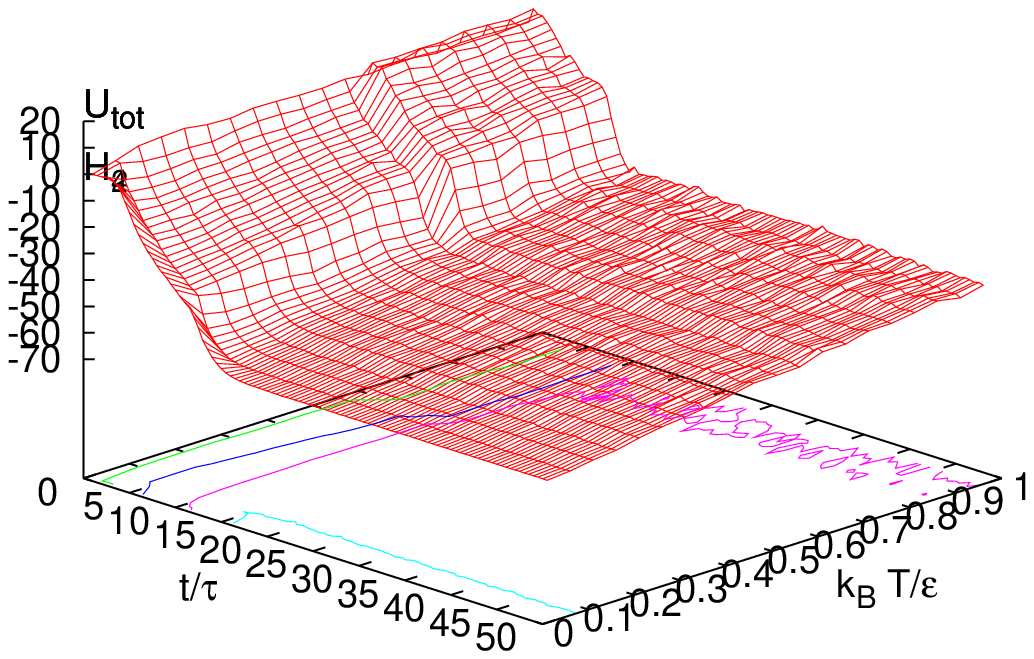}
\end{figure}

\begin{figure}[t]
\includegraphics[scale=0.7,angle=0]{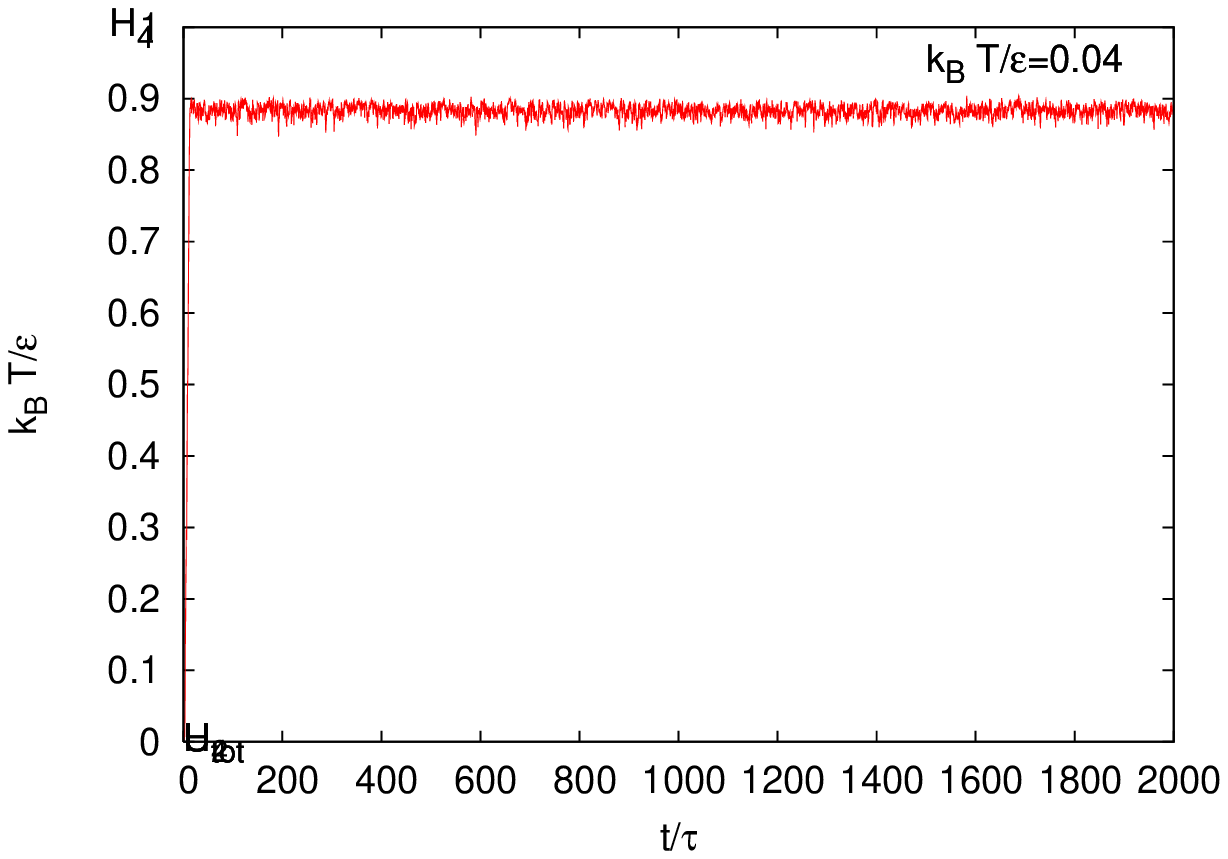}
\end{figure}



\begin{acknowledgments}
This work has been supported by the Algerian Ministry of Higher Education and
Scientific Research, within the frame work of projects CNEPRU-D01920001054.
\end{acknowledgments}

\bibliographystyle{apsrev}
\bibliography{mybib}

\clearpage

\end{document}